\documentclass[preprint,12pt,nofootinbib]{revtex4-1}

\usepackage{graphicx}\graphicspath{%
{graph/}}%
\usepackage{amssymb,amsmath,amsfonts}

\begin{document}

\begin{flushright}
{\normalsize
SLAC-PUB-16879\\
FACETII-TN-002\\
November 2016}
\end{flushright}

\title{Longitudinal Stability Study for the FACET-II e$^+$ Damping Ring \footnote[1]{Work supported by the U.S. Department of Energy under Contract No. DE-AC02-76SF00515
} }

\author{Karl Bane}
\affiliation{SLAC National Accelerator Laboratory,Menlo Park, CA 94025}

\begin{center}
\end{center}

\maketitle

\section*{Introduction}

The FACET-II e$^+$ damping ring is a small ring meant to reduce the emittances of the positron beam after the positron source. The details of the ring vacuum chamber have not yet been set. Nevertheless, we already know (see below) that shielded coherent synchrotron radiation (CSR) is a dominant impedance source. We will obtain a rudimentary ring impedance by adding to CSR the RF cavity impedance and necessary pipe transition impedances. But since these contributions are quite small compared to that of CSR, it will be difficult to accurately estimate their impact on the microwave threshold. At this stage we will aim to find the threshold approximately. In the future, when all the important vacuum chamber components are accounted for, we hope to find that at the nominal charge of $Q=1$~nC the beam is either below, or at most slightly above, the threshold to the microwave instability. 

Below we obtain a beginning impedance budget, generate pseudo-Green function wakes, and perform longitudinal phase space simulations. 
In our simulations, to make the problem manageable, we artificially reduce the damping time to $\tau_z=200T_s$, and simulate for different values of bunch charge $Q$. This is followed by changing the damping time by a factor of 2 and of 3 to verify that, locally, the results are insensitive to $\tau_z$.  But in reality the damping time is much larger, with $\tau_z=17,700T_s$. It should be verified in the future that a different phenomenon does not manifest itself in the case of such weak damping.


\section*{Impedance Budget}

\begin{figure}[h]
    \begin{center}
    \includegraphics[width=0.65\textwidth]{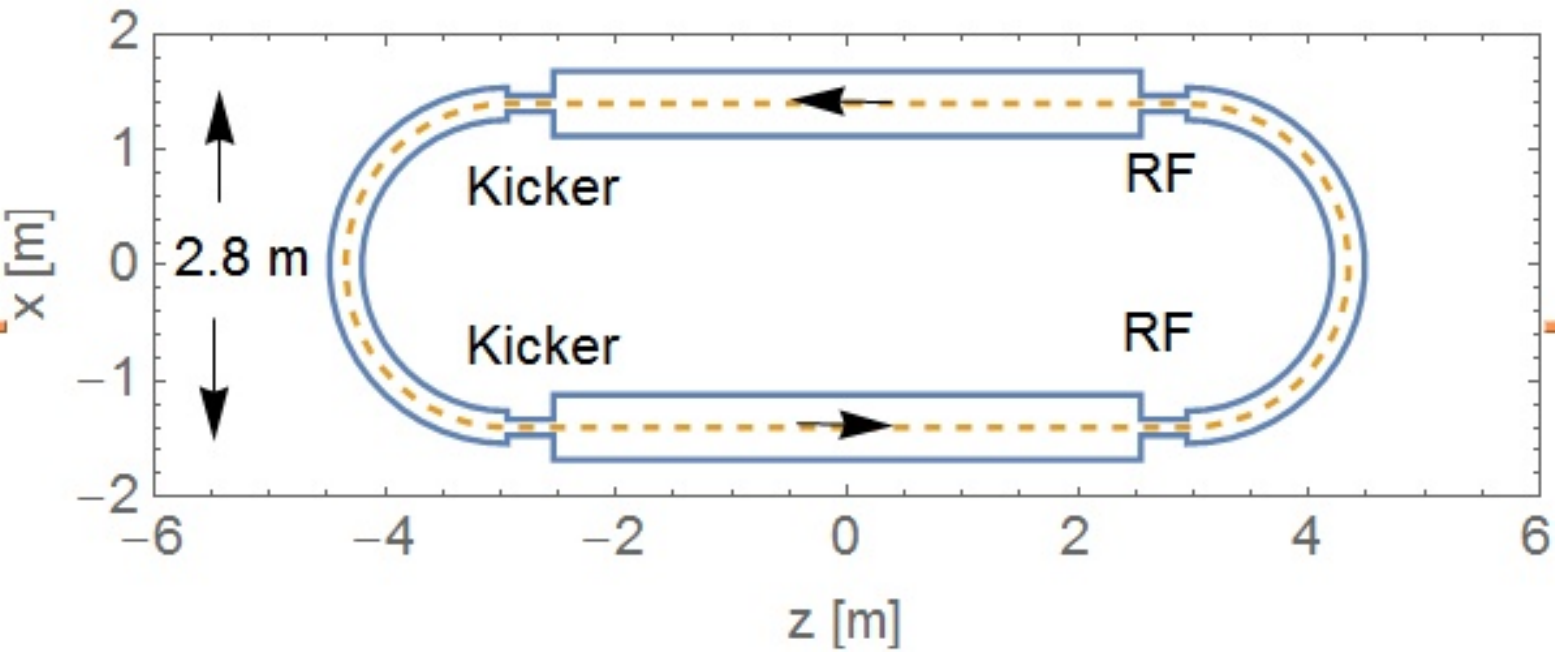}
    \caption{Simplified sketch of the layout of the FACET e$^+$ damping ring used in this study, showing the relative size of the beam pipes in the different regions. The beam pipes are round with radius $a=0.75$, 2.0, 4.0~cm in the RF cavity and kicker sections, in the arcs, and in the straight sections, respectively. The locations of the cavities and kickers are also indicated. The arrows give the beam's direction of motion.}\label{sketch_fi}
    \end{center}
    \end{figure}

To estimate the microwave threshold for the Facet-II e$^+$ ring we begin by generating an impedance budget. Having as yet no engineering drawings of the vacuum chamber we consider only: the two 2-cell RF cavities,four pairs of transitions,  the kicker chambers, the wall resistance, and the effect shielded coherent synchrotron radiation (CSR). Other objects that may be important and can be added in the future are the button beam position monitors (BPMs), the shielded bellows, the septa, etc. In general, the beam pipes of the ring are round. Fig.~\ref{sketch_fi} gives a simplified sketch of the layout of the ring showing the relative beam pipe sizes and the locations of the two RF cavities and the two kickers. The beam pipes are round with radius $a=0.75$, 2.0, 4.0~cm in the RF cavity and kicker sections, in the arcs, and in the straight sections, respectively. The large aperture in the straight sections is needed to accept the injected beam with its large emittance.

\subsection*{RF Cavities}

\begin{figure}[h]
    \begin{center}
    \includegraphics[width=0.8\textwidth]{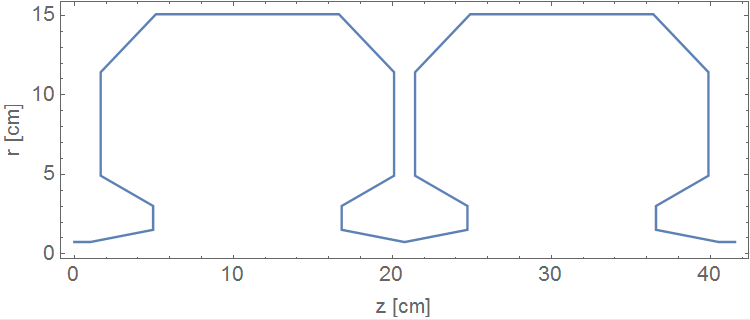}
    \caption{The shape of one 2-cell SLC damping ring RF cavity. The beam pipe radius, as well as the minimum aperture, is 7.5~mm.}\label{cavity_fi}
    \end{center}
    \end{figure}

\begin{figure}[h]
    \begin{center}
    \includegraphics[width=0.7\textwidth]{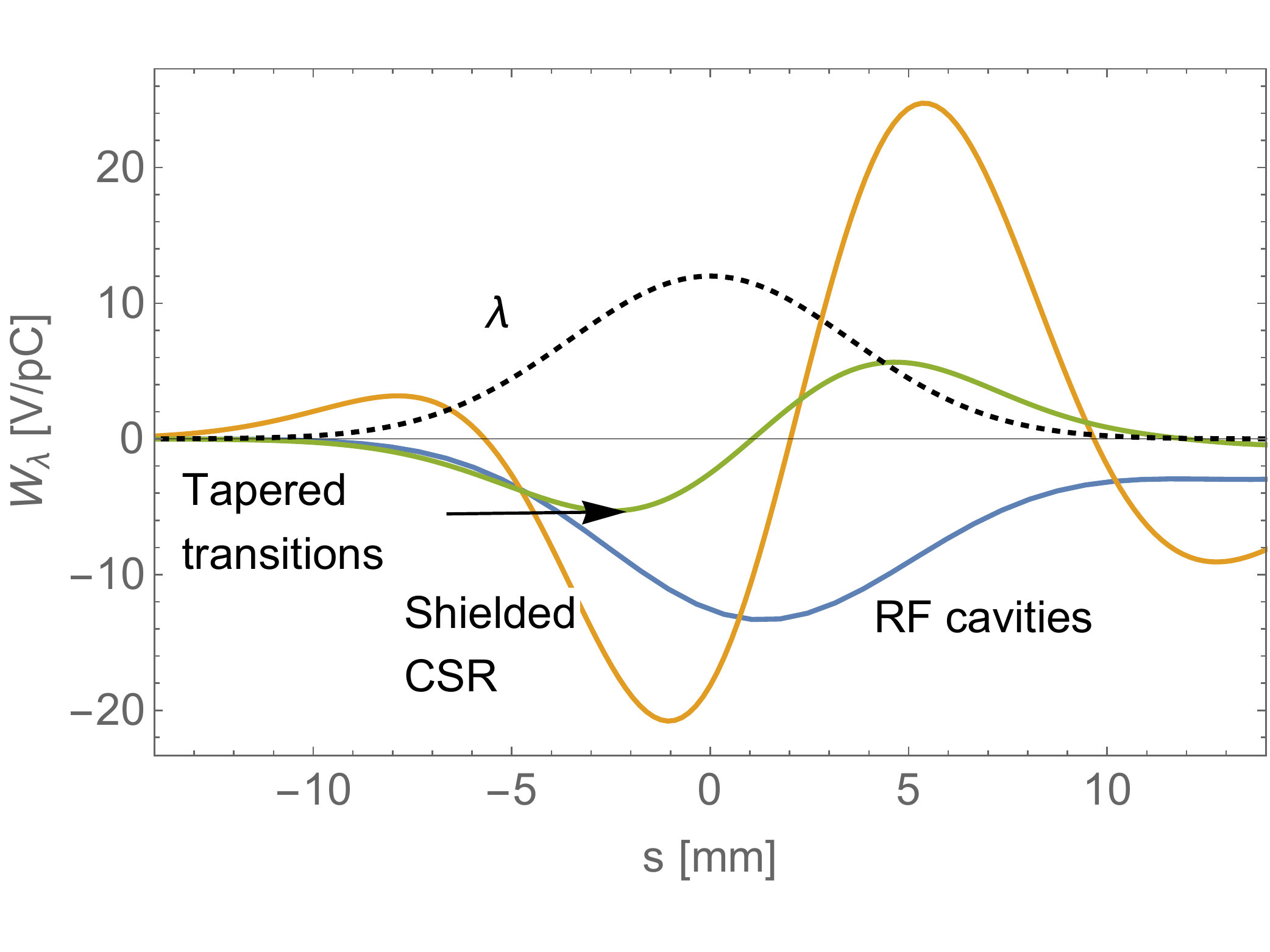}
    \caption{The bunch wake for a Gaussian bunch with $\sigma_z=3.5$~mm due to: two 2-cell RF cavities (blue curve),  the shielded CSR in the ring (red), and the 8 tapered transitions (green). The bunch distribution $\lambda(s)$, with the head to the left, is also shown (black dots).}\label{wake_comp_fi}
    \end{center}
    \end{figure}

The RF cavities are the two 2-cell, 714 MHz cavities that were used in the SLC damping rings (see Fig.~\ref{cavity_fi}). They are cylindrically symmetric, and one cavity will be located in each straight section of the FACET ring. The external beam pipes, as well as the minimum aperture, have a radius of 7.5~mm. To obtain the wakefield, we use I.~Zagorodnov's ECHO program \cite{Zag05}, using as driving bunch a Gaussian with rms length $\sigma_z=3.5$~mm (the nominal bunch after it has been lengthened by IBS). The bunch wake $W_\lambda(s)$ of the two 2-cell cavities in the ring is shown in Fig.~\ref{wake_comp_fi} (the blue curve); the bunch shape $\lambda(s)$ is also shown, with the head to the left (black dots).

The wake of the RF cavities for the SLC damping rings has been approximated with a resistive model~\cite{Bane88}; {\it i.e.} one given by a bunch wake of the form
\begin{equation}
W_{R}(s)=-cR\lambda(s)  \ ,
\end{equation}
with $c$ the speed of light and $R$ the resistance, a constant. From Fig.~\ref{wake_comp_fi} we see that the resistive model is a good approximation to the cavity bunch wake ({\it i.e.} the shape of the cavity wake is similar to $-\lambda(s)$). Fitting the bunch wake to the model (and weighting by the Gaussian distribution) we find an effective (per cavity) resistance, $R=193$~$\Omega$. 

\subsection*{Transitions}

The beam pipe in the ring is mostly round, with radius $a=2$~cm in the arcs, $a=4$~cm in the straights, and $a=0.75$~cm in the RF cavity and kicker sections (see Fig.~\ref{sketch_fi}). We see that at least 8 (round) beam pipe transitions are needed in the ring, four connecting $a=0.75$~cm to 2~cm, and four connecting $a=0.75$~cm to 4~cm. Let us first consider non-tapered step transitions. 
If $\sigma_z/a_1\ll1$, where $a_1$ is the smaller of the two beam pipes (which is approximately satisfied here, where $\sigma_z=3.5$~mm and $a_1=7.5$~mm), we can use the optical model to approximate the wake of the step transitions. According to this model, the wake of a symmetric pair of step transitions (between radius $a_1$ and $a_2$ with $a_2>a_1$) is resistive with resistance
given by~\cite{Handbook} 
\begin{equation}
R=\frac{Z_0}{\pi}\ln\left(\frac{a_2}{a_1}\right)\ \quad\quad \quad\quad (a_2>a_1)\ ,
\end{equation}
with $Z_0=377$~$\Omega$. If we add the effect of all 8 transitions (four transition pairs), we find that $R_{tot}\approx637$~$\Omega$. Thus, the beam pipe transitions in the FACET ring---if not tapered---have the same type of impedance, but a factor 1.6 times stronger, than the two RF cavities.

We can reduce the strength of the impedance of the transitions by tapering them. 
However, we cannot taper them too much, since we are limited in how much of the ring circumference is available for this purpose.
When transitions are sufficiently tapered, their impedance becomes primarily inductive, and the more the taper, the smaller the impedance becomes. We choose to taper the transitions with taper angle $\theta=10^\circ$. With the 8 transitions, this means that the tapers---in total---will take up 1.0~m of the ring circumference.


For a symmetric pair of tapered transitions connecting beam pipes of radii $a_1$ and $a_2$ ($a_2>a_1$) at a small angle $\theta$, the impedance is approximately inductive, $Z(\omega)=i\omega{\cal L}$, with $\cal L$ the inductance, where~\cite{Yokoya}
\begin{equation}
{\cal L}=\frac{Z_0}{2\pi c}(a_2-a_1)\tan\theta\quad\quad\quad\quad\quad(a_1\tan\theta\ll c/\omega,\ a_2>a_1)\ .
\end{equation} 
For this estimate to be valid requires that, for our Gaussian bunch, $\sigma_z\gg a_1\tan\theta$ (for $a_1=0.75$~cm, this means that $\theta\ll 25^\circ$). Note that an important parameter in instability theory is $|{\bar Z/n}|$, with $\bar Z$ the impedance at a representative frequency, and $n=\omega/\omega_0$, were $\omega$ is frequency and $\omega_0=2\pi c/C$, with $C$ the ring circumference. For an inductive impedance $|{\bar Z/n}|=\omega_0{\cal L}$. 

We have performed ECHO calculations for the nominal Gaussian bunch, for a pair of both types of transitions. The sum of the wakes representing all 8 transitions is given by the green curve in Fig.~\ref{wake_comp_fi}. We have assumed that the transitions are not close to one another or close to the cavities (we assume their spacing is at least several times the larger pipe radius). We see that the total wake is smaller in amplitude than that of the RF cavities (blue curve), and that the character is markedly inductive---{\it i.e.} the shape is similar to $-\lambda(s)'$. We fit the wakes to the sum of a purely resistive wake plus a purely inductive wake:
\begin{equation}
W_{R+L}(s)=-cR\lambda(s) -c^2{\cal L}\lambda'(s)  \ .
\end{equation}
Per transition we find that the fitted $R=3.3$~$\Omega$ and ${\cal L}=0.22$~nH for the 0.75--2~cm transitions, and $R=10.6$~$\Omega$ and ${\cal L}=0.05$~nH for the 0.75--4~cm transitions.
We define a goodness-of-fit parameter by
\begin{equation}
g_{fit}=1-\sqrt{{\int[W_\lambda(s)-W_{R+L}(s)]^2\lambda(s)\,ds}}\Bigg/\sqrt{{\int W^2_\lambda(s)\lambda(s)\,ds}}\ .
\end{equation}
For both of the transitions we find that the model fits well, and that $g_{fit}=0.74$.

We can approximate $|{\bar Z/n}|$ from the model fit using
\begin{equation}
\frac{{\bar Z}}{n}=\frac{\omega_0\sigma_z}{c}R+i\omega_0{\cal L}\ .
\end{equation}
We find that for the four 0.75--2~cm transitions, $|{\bar Z}/n|=0.04$~$\Omega$; for the four 0.75--4~cm transitions, $|{\bar Z}/n|=0.18$~$\Omega$.

\subsection*{Shielded CSR Wake}

When the FACET bunch moves through the arcs of the ring it will radiate energy coherently, and the effect on the beam can be described by the parallel plate model of coherent synchrotron radiation (CSR)~\cite{MGK}. 
The model considers the CSR wakefield generated by an electron moving on a
circular orbit with bending radius $\rho$ in the middle of two
parallel plates separated by distance $2h$. This steady-state formula is applicable when the so-called { formation length}, $\ell_f=4a^2/(\pi^2\sigma_z)$, is small compared to the magnet lengths~\cite{Derbenev}. Taking $a=2$~cm, $\sigma_z=3.5$~mm, we find that $\ell_f=5$~cm, which is indeed small compared to the magnet lengths; thus the steady-state formula for CSR does apply.

In the case of no shielding the wake is non-zero only for negative $s$, ({\it i.e.} with the test particle ahead of the driving charge). The point charge wake is given by
\begin{equation}
W_{\delta0}(s)=-\frac{Z_0c}{3^{4/3}}H(-s)\frac{\rho^{1/3}}{(-s)^{4/3}}\ ,
\end{equation}
with $H(s)=1$ for $s\ge0$, $H(s)=0$ for $s<0$.
With shielding, the wake $W_{\delta}(s)=W_{\delta0}(s)+W_{\delta1}(s)$, with
\begin{equation}
W_{\delta1}(s) =  -\Big(\frac{Z_0c}{4\pi}\Big)\rho^{1/3}\Big({\Pi\over \sigma_{z0}}\Big)^{4/3}G(-\frac{\Pi s}{\sigma_{z0}}), \label{eqn:wake}
\end{equation}
where the shielding parameter $\Pi=\sigma_{z0}\rho^{1/2}/h^{3/2}$, and $\sigma_{z0}$ is the nominal bunch length. The term $W_{\delta1}(s)$ is in general non-zero for both signs of argument. The function $G(\zeta)$ is given by
 \begin{equation}
 G(\zeta) = 8\pi\sum_{k=1}^\infty {(-1)^{k+1}\over k^2} {Y_k(\zeta)[3-Y_k(\zeta)]\over [1+Y_k(\zeta)]^3},
 \end{equation}
where $Y_k$ is a root of the equation
\begin{equation}
Y_k - {3\zeta\over k^{3/2}} Y_k^{1/4} - 3 = 0.
\end{equation}
This equation has two real, positive roots and two complex roots. We choose the smaller real root when $\zeta<0$, the larger real root when $\zeta>0$. Normally we sum $k$ up to 25.

To find the bunch wake we convolve the two pieces of the point charge wake with the bunch distribution (the first piece is handled specially), and then add the results. In detail, we find
\begin{equation}
W_{\lambda0}(s)=-\frac{Z_0c}{3^{1/3}}\rho^{1/3}\int_0^s\frac{\lambda'(s-s')}{(s')^{1/3}}\,ds'\ , \quad\quad W_{\lambda1}(s)=\int_{-\infty}^\infty W_{\delta1}(s')\lambda(s-s')\,ds'\ ,\label{convolve_eq}
\end{equation}
where $\lambda'(s)\equiv d\lambda(s)/ds$.

In the FACET damping ring the beam pipe in the arcs is round, with radius $a=2$~cm. We take this to be the half-gap between the two plates of the model, {\it i.e.} we let $h=2$~cm. In addition, $\rho=0.78$~m and the nominal bunch length (including IBS) $\sigma_{z0}=3.5$~mm. The wake of a Gaussian bunch due to shielded CSR is shown by the red curve in Fig.~\ref{wake_comp_fi}. We see that the CSR wake, at the nominal bunch length, dominates over the contributions of the RF cavities and of the tapered transitions. Fitting to the model of a resistive plus inductive wake we obtain $R=373$~$\Omega$ and ${\cal L}=7.4$~nH, and a goodness of fit parameter, $g_{fit}=0.43$. We see that the model does not fit well. 

\begin{figure}[h]
    \begin{center}
    \includegraphics[width=0.7\textwidth]{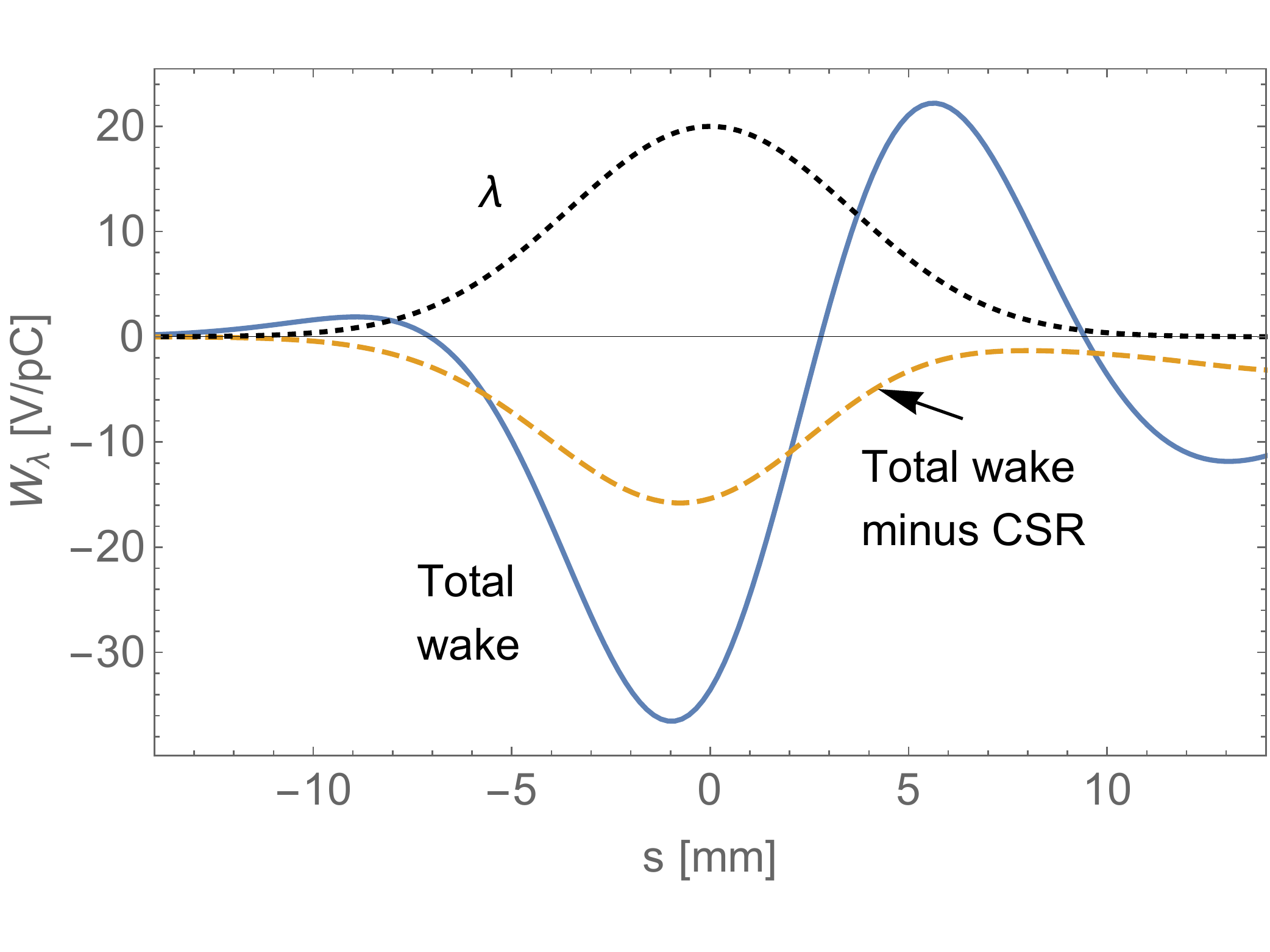}
    \caption{The total, single turn, bunch wake for a Gaussian bunch with $\sigma_z=3.5$~mm passing through the storage ring  (assuming tapered transitions; the blue curve). The red dashed curve gives the total wake minus the shielded CSR contribution. The bunch distribution $\lambda$, with the head to the left, is also shown (black dots).}\label{wake_tot_fi}
    \end{center}
    \end{figure}

In Fig.~\ref{wake_tot_fi} we give the total wake---the sum of RF cavity, transition, and CSR contributions---for the $3.5$~mm Gaussian bunch (the blue curve). In addition, we plot the wake minus the CSR contribution in dashed red, and see that CSR dominates the wake. 

In Table~\ref{imp_tab} we give an impedance budget, summarizing the fitting discussed above. We also include the effect of the two kicker chambers and the wall resistance, both of which contribute a small amount of impedance compared to the three types of wake sources discussed above. In total we have resistance $R=832$~$\Omega$, inductance ${\cal L}=10.4$~nH, and effective impedance factor $|{\bar Z}/n|=1.45$~$\Omega$. Finally, note that if we take the total wake function of Fig.~\ref{wake_tot_fi} (the blue curve) and fit to the $R+{\cal L}$ model, we see that there is some cancellation: $R=823$~$\Omega$, ${\cal L}=8.8$~nH, and  $|{\bar Z}/n|=1.19$~$\Omega$. Here the goodness-of-fit parameter $g_{fit}=0.83$.

\begin{table}[h!]
\centering
\caption{Simple impedance budget for the Facet-II e$^+$ damping ring. To quantify the contribution of the objects, the wake of a Gaussian bunch with $\sigma_z=3.5$~mm, obtained by ECHO, was fit to that of the sum of a purely resistive plus inductive object, with the fit weighted by the Gaussian bunch distribution (the RF cavity wake was fit to only a resistive impedance). Column 4 gives the goodness of fit, $g_{fit}$. The effective $|{\bar Z}/n|$ is obtained from the fitted $R$ and $\cal L$.}
\begin{tabular}{||l||c|c|c||c|c|c|c||} \hline\hline
 & \multicolumn{3}{|c||}{Single Element}&\multicolumn{4}{|c||}{Total Contribution}\\ \cline{2-8}
 \raisebox{2.5ex}[0pt]{\hspace{10mm}Object}& $R$ [$\Omega$] & $\cal L$ [{nH}] & $g_{fit}$ & { Number}& $R$ [$\Omega$] & $\cal L$ [{nH}] & $|{\bar Z}/n|$ [$\Omega$]\\ \hline\hline
RF cavities & 193. &--- & 0.78& 2 & 386. &--- &0.41\\ \hline  
Tapered transitions: &  &  & &   &  &  &\\ \cline{2-8} 
\hspace{5mm}0.75--2 cm & 3.3 & 0.22 &0.74 & 4 &13.2 & 0.86& 0.04\\ \cline{2-8}
\hspace{5mm}0.75--4 cm & 10.6 & 0.50 &0.74 &4 &42.6 & 1.98 & 0.18\\ \hline 
Kicker chambers & 7. &0.04 &0.57 & 2 & 14. &0.08& 0.02\\ \hline 
Resistive wall & 3.5 & 0.05& 0.84& 1 & 3.5 &0.05 & 0.01\\ \hline
Shielded CSR & 373. &7.4 & 0.43& 1 & 373. &7.4 & 0.79\\ \hline\hline 
Total & & & & &832. &10.4 &1.45 \\ \hline\hline
\end{tabular}\label{imp_tab}
\end{table}

\section*{Instability}

\subsection*{Boussard Criterion}

As a first estimate of the instability threshold, often the Boussard criterion is used. It is known to give a rough and conservative estimate of the threshold to a strong instability. According to this criterion, the threshold bunch charge is given by \cite{Boussard72}
\begin{equation}
Q_{th}=(2\pi)^{3/2}\frac{\alpha\sigma_z E\sigma_\delta^2}{c|{\bar Z}/n|}\ ,
\end{equation}
with $\alpha$ the momentum compaction factor, $E$ the beam energy, $\sigma_\delta$ the relative beam energy spread, and ${\bar Z}/n$ the impedance at a representative frequency. For the FACET ring, taking $\alpha=0.0584$, $E=335$~MeV, $\sigma_\delta=7.2\times10^{-4}$ (IBS increased value), and $|{\bar Z}/n|=1.19$~$\Omega$ (the effective impedance obtained from the final wake), we obtain $Q_{th}=1.59$~nC, which is 60\% larger than the design charge.

Besides the strong instability, there is also the so-called {\it weak instability}~\cite{Oide}. It is weak in that it can be Landau damped by incoherent tune spread, and the threshold varies with the longitudinal damping time as $Q_{th}\sim \tau_z^{-1/2}$. If the impedance is resistive, which implies it has little tune spread, then the weak instability can be excited at a lower current than the strong instability. 

\subsection*{CSR Instability}

It can be shown that for a beam in a machine dominated by the CSR impedance, the instability threshold depends only on two dimensionless parameters, the shielding parameter $\Pi$ and and the normalized charge $S_\mathrm{csr} = I\rho^{1/3} / \sigma_{z0}^{4/3}$, where 
\begin{equation}
I=\left(\frac{Z_0c}{4\pi}\right)\frac{eQ}{2\pi\nu_{s0}E_0\sigma_{\delta0}}\ ,
\end{equation}
with $Q$ the bunch charge; with $\nu_{s0}$, $E_0$, $\sigma_{\delta0}$, the nominal (zero current) values of synchrotron tune, energy, and relative energy spread~\cite{BCS}.
From simulations of the Vlasov-Fokker-Planck equation, it was found that, for a ring consisting of the CSR impedance alone, the threshold to the microwave instability is well approximated by $(S_\mathrm{CSR})_{\mathrm{th}}=0.50+0.12\Pi$~\cite{BCS}, a result that is true away from the region $\Pi\sim0.35$--1.0. (Within this region, the  {weak instability} appears already at a lower current, one whose exact value depends on the longitudinal damping time.)
Note that for some storage rings with short bunches, such as the Metrology Light Source~\cite{MLS} and ANKA~\cite{ANKA} in Germany, this simple model has been shown to agree quite well with threshold measurements. 

In the FACET damping ring the beam pipe in the arcs is round, with radius $a=2$~cm. We take this to be the half-gap between the two plates of the model, {\it i.e.} we let $h=2$~cm. In addition, $\rho=0.78$~m, $\nu_{s0}=0.0383$, $E_0=335$~MeV;  $\sigma_{z0}=3.5$~mm and $\sigma_{\delta0}=7.3\times10^{-4}$ (the relative bunch length and energy spread when including the effect of IBS). Thus, the shielding parameter $\Pi=1.11$, and the instability threshold, assuming the only impedance in the ring is the CSR impedance, is $(S_\mathrm{CSR})_{\mathrm{th}}=0.50+0.12\Pi=0.63$, which is equivalent  to a threshold charge of $Q_{\mathrm {th}}=2.4$~nC.

\subsection*{Pseudo-Green Function Wake}

For our instability simulations we need a wake function that we can convolve with a bunch distribution to obtain the bunch wake. For this purpose we numerically generate a wake for a short Gaussian driving bunch that we call the {\it  pseudo-Green function wake}, $W_\delta(s)$. 
The nominal bunch length is $\sigma_{z0}=3.5$~mm. We use ECHO to obtain the wakes of the RF cavities and transitions, using a Gaussian driving bunch with $\sigma_z=0.35$~mm, and having the calculated wakes reach to 35~mm behind the driving bunch. A similar calculation is performed analytically for the shielded CSR wake. These wakes are then summed to obtain one pseudo-Green function wake, $W_\delta(s)$, representing the entire ring. We repeat the process for the cases: (a)~shielded CSR alone, (b)~shielded CSR plus the rf cavities plus tapered transitions, and (c)~shielded CSR plus the rf cavities plus rectangular transitions. 

\begin{figure}[h]
    \begin{center}
    \includegraphics[width=0.7\textwidth]{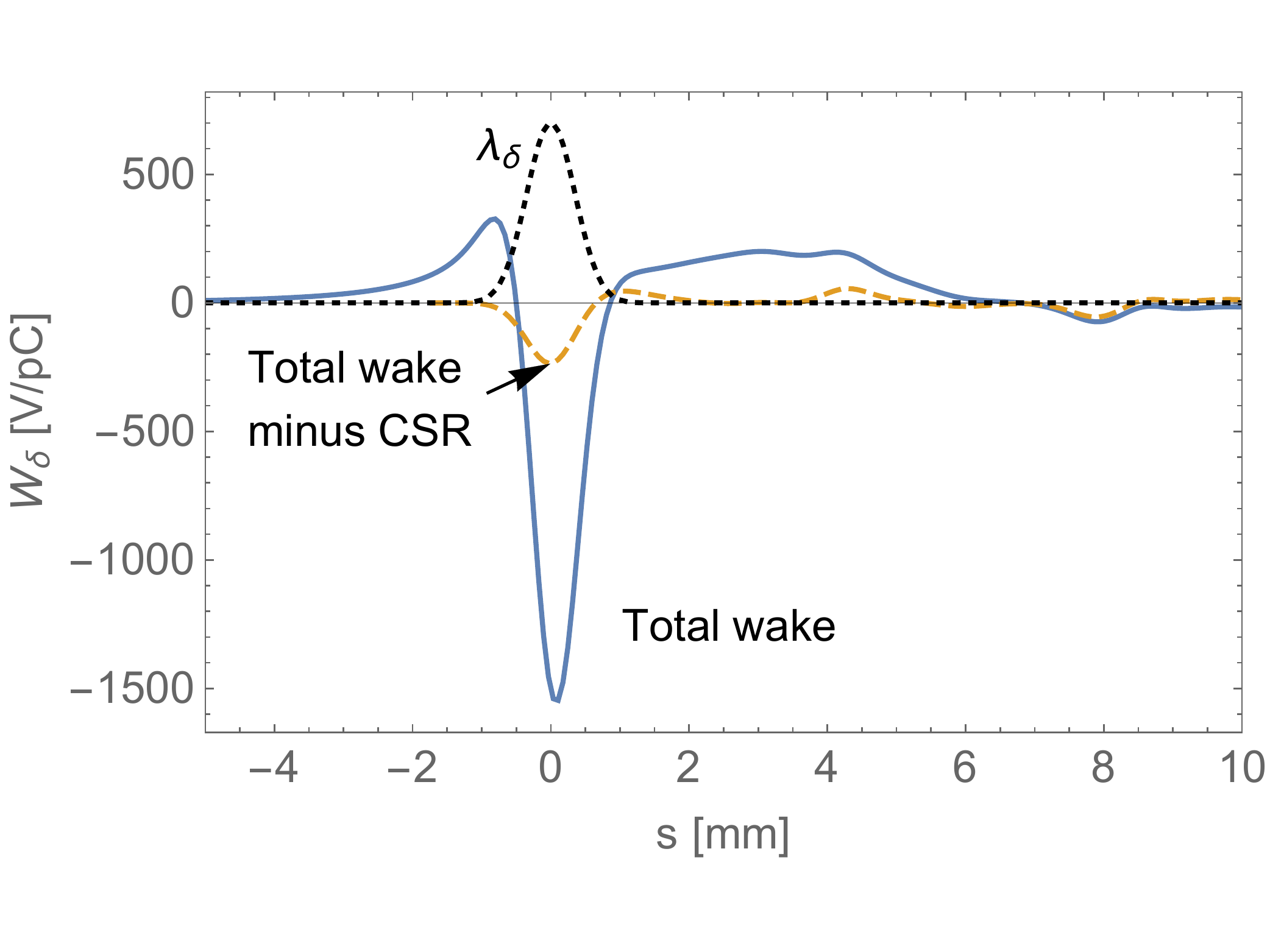}
    \caption{The pseudo-Green function wake representing the entire ring (assuming tapered transitions; blue curve), and the wake minus the CSR contribution (red dashes). The bunch distribution $\lambda_\delta(s)$, a Gaussian with $\sigma_z=0.35$~mm (with the head to the left), is also shown (black dots).}\label{Green_fcn_fi}
    \end{center}
    \end{figure}

The pseudo-Green function wake for case (b), with the tapered transitions, is shown in 
Fig.~\ref{Green_fcn_fi} (the blue curve). The dashed, red curve in the figure gives the total wake minus the CSR contribution for comparison; we see that the CSR wake greatly dominates the Green function wake. 

\subsection*{Simulations}

To study the effect on the microwave threshold of the different impedance models we use a macro-particle tracking program~(see {\it e.g.} \cite{Wilson}). For a large number $M$ of macro-particles, the tracking program follows the energy deviation normalized to $\sigma_{\delta0}$, $p_i$, and the longitudinal position normalized to $\sigma_{z0}$, $q_i$, over many time steps $\theta=2\pi\Delta t/T_s$, where $T_s$ is the synchrotron period.
The equations solved are
\begin{equation}
\left(
\begin{array}{c}
p_i \\
q_i
\end{array}
\right)_{n+1}\!\!\!\!\!\!=
\left[
\begin{array}{cc}
1-c_d & -\theta \\
(1-c_d)\theta & 1-\theta^2
\end{array}
\right]
\left(
\begin{array}{c}
p_i \\
q_i
\end{array}
\right)_{n}
+
\left[
\begin{array}{c}
1\\
\theta
\end{array}
\right] 
\left(\theta IW_\lambda(q_i\sigma_{z0})+r_e\right)\ ,
\end{equation}
with $c_d$ and $r_e$ representing effects of radiation damping and quantum excitation.
After each time step the particles are binned and the bunch distribution $\lambda(s)$ is recomputed. This is followed by performing convolutions like the right equation of Eq.~\ref{convolve_eq}, using the pseudo-Green function wake, to obtain the bunch wake.

Runs were performed for different values of charge $Q$, to find the charge at which the energy spread of the beam begins to increase from the nominal value. (Here the nominal energy spread and bunch length are the IBS increased values.) For the simulations, the number of macro-particles $M=10^5$, size of domain in $s$ is $12\sigma_{z}$, number of $s$ bins is 240, time step $\Delta t=0.02T_s$, $\tau_z/T_s=200$. We begin with the Haissinski solution for bunch distribution, $\lambda(s)$. Calculations were performed for one or two damping times. The value of $\sigma_\delta$ taken as the result was the average value over several oscillations near the end of the run. 

\begin{figure}[h]
    \begin{center}
    \includegraphics[width=0.65\textwidth]{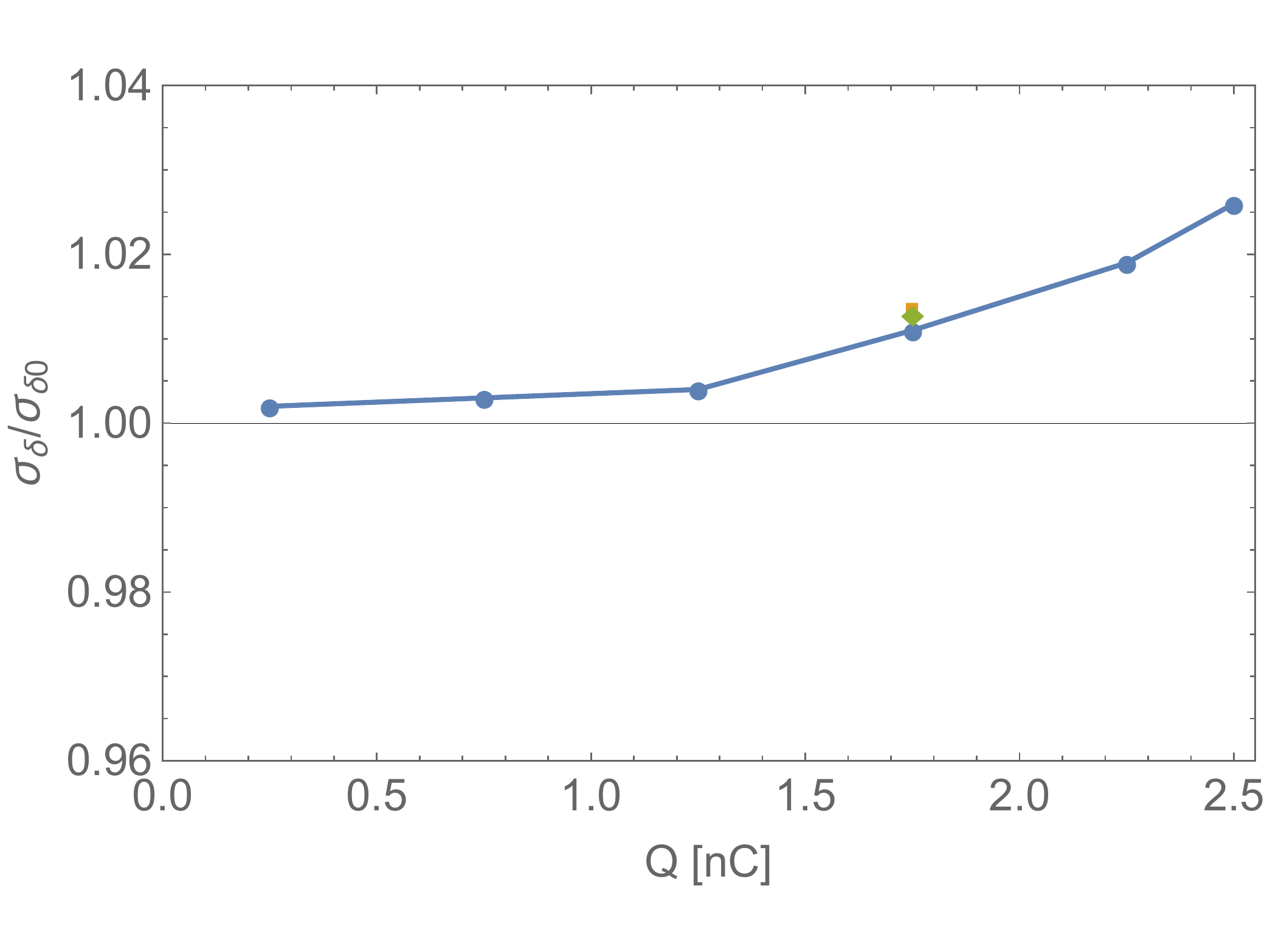}
    \caption{Simulated energy spread after two damping times for case~(b) as function of charge $Q$ (blue). At $Q=1.75$~nC, the case when the damping time was increased by a factor of 2 (3) is indicated by the red (green) symbol.
}\label{sige_totxc_fi}
    \end{center}
    \end{figure}

\begin{figure}[h]
    \begin{center}
    \includegraphics[width=0.49\textwidth]{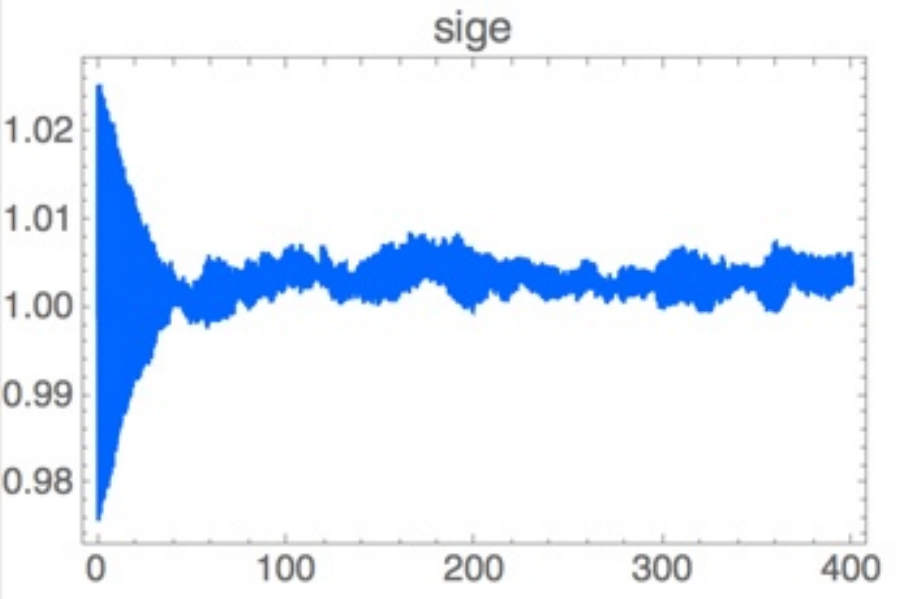}
    \includegraphics[width=0.49\textwidth]{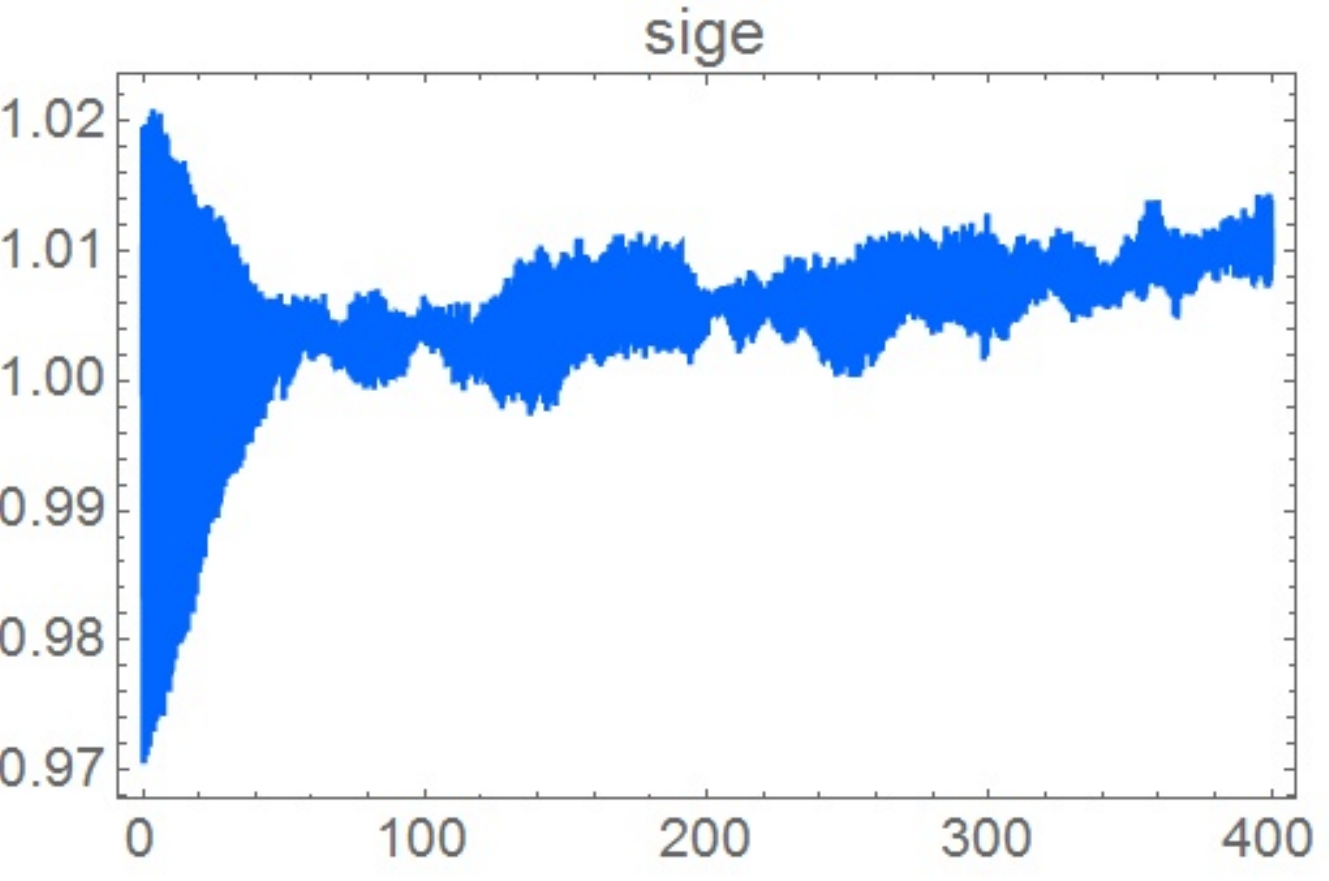}
    \caption{For case (b), shielded CSR plus RF cavities plus tapered transitions): $\sigma_\delta/\sigma_{\delta0}$ {\it vs.} $ct/\lambda_s$, for $Q=1.25$~nC (left plot) and $Q=1.75$~nC (right). }\label{sige_totxcb_fi}
    \end{center}
    \end{figure}

To get an idea of the accuracy of our calculations, we began the simulations by using as pseudo-Green function the shielded CSR wake alone, for a Gaussian bunch with $\sigma_z=0.35$~mm, and varying $Q$ in steps of 0.5~nC (case a). We obtained a rather smooth curve of $\sigma_\delta$ {\it vs.} $Q$, with the threshold at $Q_{th}\sim1.75$--2.25~nC. But from the analytical fit of the (more careful) simulations of Ref.~\cite{BCS}, discussed earlier, $Q_{th}=2.4$~nC. Thus, believing the earlier results to be correct, we conclude that the current simulations have a slight error, one that is pessimistic. Next we repeated the calculation for case (b), shielded CSR plus RF cavities plus tapered transitions. The resulting $\sigma_\delta$ {\it vs.} $Q$ curve is shown in Fig.~\ref{sige_totxc_fi}. The threshold appears to be at $Q_{th}=1.25$~nC. In Fig.~\ref{sige_totxcb_fi} we give more details, showing the development of $\sigma_\delta/\sigma_{\delta0}$ {\it vs.} $t/T_s$, for $Q=1.25$~nC and 1.75~nC. We see that, after an initial damping of oscillations, the results of the former case are flat with time, whereas those of the latter case have an increasing slope. In Fig.~\ref{sige_totxc_fi}, with the red (green) plotting symbol, we show the simulation result for $Q=1.75$~nC when the damping time was increased by a factor of 2 (3). One can see that the energy spread is not locally sensitive to damping time. Finally, we repeated the calculations for case (c), shielded CSR plus RF cavities plus rectangular transitions. The result is $Q_{th}=1.00$--1.25~nC. All the simulation results are summarized in Table~\ref{Qth_tab}.

\begin{table}[h!]
\centering
\caption{Threshold to the microwave instability in the FACET-II e$^+$ damping ring obtained by macro-particle simulation under various assumptions for the impedance. According to the more accurate calculations of Ref.~\cite{BCS} the result for (a) should be $Q_{th}=2.4$~nC.}
\begin{tabular}{||c|l||c||} \hline\hline
Case&Ring Impedance Model & $Q_{th}$ [nC]\\ \hline\hline 
(a)&Shielded CSR only & 1.75--2.25\\ \hline
(b)&Shielded CSR + RF cavities + tapered transitions& 1.25\\ \hline
(c)&Shielded CSR + RF cavities + rectangular transitions&1.00--1.25\\ \hline\hline
\end{tabular}\label{Qth_tab}
\end{table}

\section*{Conclusions}

We have performed initial studies of the microwave instability in the FACET-II e$^+$ damping ring, considering a rudimentary collection of impedance objects: shielded CSR, the RF cavities, and beam pipe transitions. For the case where the transitions are tapered to angle $\theta=10^\circ$, we find the total effective resistance, inductance, and impedance: $R=823$~$\Omega$, ${\cal L}=8.8$~nH, and  $|{\bar Z}/n|=1.19$~$\Omega$. Shielded CSR dominates the impedance. 

We have performed macro-particle simulations using the pseudo-Green function approach and find that the ring is stable at the nominal charge $Q=1$~nC with these three kinds of objects, whether the transitions are tapered or not. The fact that CSR dominates the impedance is part of the reason why it is difficult to accurately find the impact of the other objects. We have indications that the results for the threshold given here are conservative. 

The simple (parallel plate) CSR model used here has worked well in predicting the bursting threshold in short bunch machines like MLS and ANKA in Germany. Nevertheless, since the CSR contribution dominates here, it might be worthwhile putting effort into CSR calculations using more realistic models---for example R. Warnock~\cite{BobMarit} and Y. Cai~\cite{Yunhai} have programs that, in principle, can calculate CSR in a square beam pipe. Also, as they become available, other impedance objects need to be included in the studies, especially objects---like the beam position monitor---that are numerous and likely significant. 

In our simulations we artificially reduced the longitudinal damping time $\tau_z$ by a factor of 100, to make the study tractable. We checked that the results were insensitive locally to changes in $\tau_z$ (to changes by a factor of 3). Nevertheless, it would still be prudent to study and verify that, with the much longer damping time of the real machine, no new physics manifests itself and invalidates the conclusions of this report. 

\section*{Acknowledgments}
The author thanks M. Woodley and Y. Cai for helpful discussions about this ring.

\end{document}